\documentclass[a4paper,11pt]{article}
\usepackage{pos}
\usepackage{euclid}
\usepackage{natbib}
\usepackage{amsmath}

\title{Construction of spectral libraries of star-forming galaxies at $0.3\leq z \leq 2.5$}

\author*{Louis Gabarra\textsuperscript{\normalfont\textit{a,b}}, on behalf of the Euclid Consortium}
\affiliation[a]{ Dipartimento di Fisica e Astronomia ”G. Galilei”,\\ Università di Padova, 35131, Padua, Italy}
\affiliation[b]{INFN, Istituto Nazionale di Fisica Nucleare,\\ Sezione di Padova, 35131 Padua, Italy}

\emailAdd{louis.gabarra@pd.infn.it}

\abstract{Spectral energy distribution (SED) models of galaxies at $0.3\leq z \leq 2.5$ were constructed to be processed by the \Euclid spectroscopic channel simulator in order to investigate the \Euclid Near-Infrared Spectrometer and Photometer (NISP) capabilities.

We developed a solid methodology to build a realistic and representative synthetic SED library of star-forming galaxies,  including reliable emission line fluxes and widths. The construction of the SEDs consists in computing the continuum using the Bruzual \& Charlot (2003) models, calling out SED fitting parameters available in publicly released multi-wavelength catalogues. Emission lines were added using empirical and theoretical relations.}
\FullConference{%
  41st International Conference on High Energy physics - ICHEP2022\\
  6-13 July, 2022\\
  Bologna, Italy
}

\begin{document}
\maketitle

   
\section{Context}
\Euclid will provide tens of millions of spectra in a redshift range yet under-sampled. This redshift range includes the cosmic noon at $z\approx 2$, when galaxies were particularly prolific in producing stars. Therefore, beyond the primary cosmological science of the mission, an unprecedented amount of data will be available to revolutionise galaxy evolution study. The analysis will provide crucial information through line diagnosis and scaling relations. In order to be prepared when the huge amount of data will be released, an efficient strategy to fine tune the data reduction and processing pipelines is to work with simulations. We present in this work a robust methodology to build a spectral library of star-forming galaxies (SFGs) at $0.3\leq z \leq 2.5$ that can be processed through the \Euclid spectroscopic channel simulator. 

\section{Method}
\subsection{Starting from publicly released photometric catalogues}

To construct the spectral library, we started from publicly released catalogues from the Cosmic Assembly Near-infrared Deep Extragalactic Legacy Survey (CANDELS), namely the BARRO2019 catalogue (\cite{B19}; B19 hereafter) covering the Great Observatories Origins Deep Survey North field (GOODS-N) and the COSMOS2015 catalogue (\cite{L16}; L16 hereafter) covering the COSMOS field. Both catalogues are well furnished with multi-wavelength observation from ground and space-based telescopes. The coverage from the UV to the far infrared allowed inferring robust SED fitting parameters using the Lephare software, e.g, $z$, star-formation rate (SFR), total stellar mass ($M_\star$). We selected sources within the NISP photometric sensitivity and identified as SFGs using colour-colour diagrams (see \cite{gabarrathesi} and Euclid Collaboration: Gabarra et al., in prep., for further details). We present the selected sources from the COSMOS2015 catalogue in Table \ref{tabular:GOODSN_16938}.
\begin{center}
\captionof{table}{Description of the SED-fitting parameters of the 24\,259 sources selected from the L16 catalogue that covers the COSMOS field. The selection criteria account for the redshift range of interest, the magnitude limit of the NISP photometric channel, i.e.,  $H_{\rm E} \leq 24\,\text{mag}$, and on the star-formation activity following colour-colour diagrams (see Euclid Collaboration: Gabarra et al., in prep).}
\begin{tabular}{ c|c|c|c|c|c|c}
\hline
  \multicolumn{7}{c}{\textsf{\textbf{Sources selected from the COSMOS2015 catalogue covering the COSMOS field}}}\\
\hline
 Redshift & Number & $H$  & \text{$\logten$}($M_{\star}$) 
&\text{$\logten$}(\text{SFR})   & \text{$\logten$}(Age)&   R$_{\rm e}$
\\
bins&sources&[AB mag]&[$M_{\odot}$]&[$M_\odot\,{\rm yr}^{-1}$])& [yr]&[arcsec]\\
\hline\hline
$0.3\leq z \leq0.5$ & 7356  & 17--23.5   & 8.0--11.7 & $-$19.91--2.28 & 8.0--10.0 &  0.01--14.3  \\ 
$0.5< z \leq1.0$    & 14\,603  & 17.3--24   & 8.6--11.7 & $-$19.49--2.83 & 8.0--9.9  & 0.02--13.4    \\ 
$1.0< z \leq1.5$    & 1830   & 18.7--23.6 & 8.8--11.8 & $-$10.35--2.95 & 8.0--9.7  &       0.03--6.3    \\
$1.5< z \leq2.0$    & 362    & 19.8--23.8 & 9.4--11.5 & $-$9.58--2.98  & 8.0--9.6  &  0.0--12.8    \\
$2.0< z \leq2.5$    & 108   & 20.5--23.5 & 9.4--11.5 & $-$3.01--2.71  & 8.0--9.4  &       0.1--14.2   \\
\hline
\end{tabular}

     . 
\label{tabular:GOODSN_16938}
\end{center}
\begin{figure}[ht]
    \includegraphics[width=0.93\textwidth]{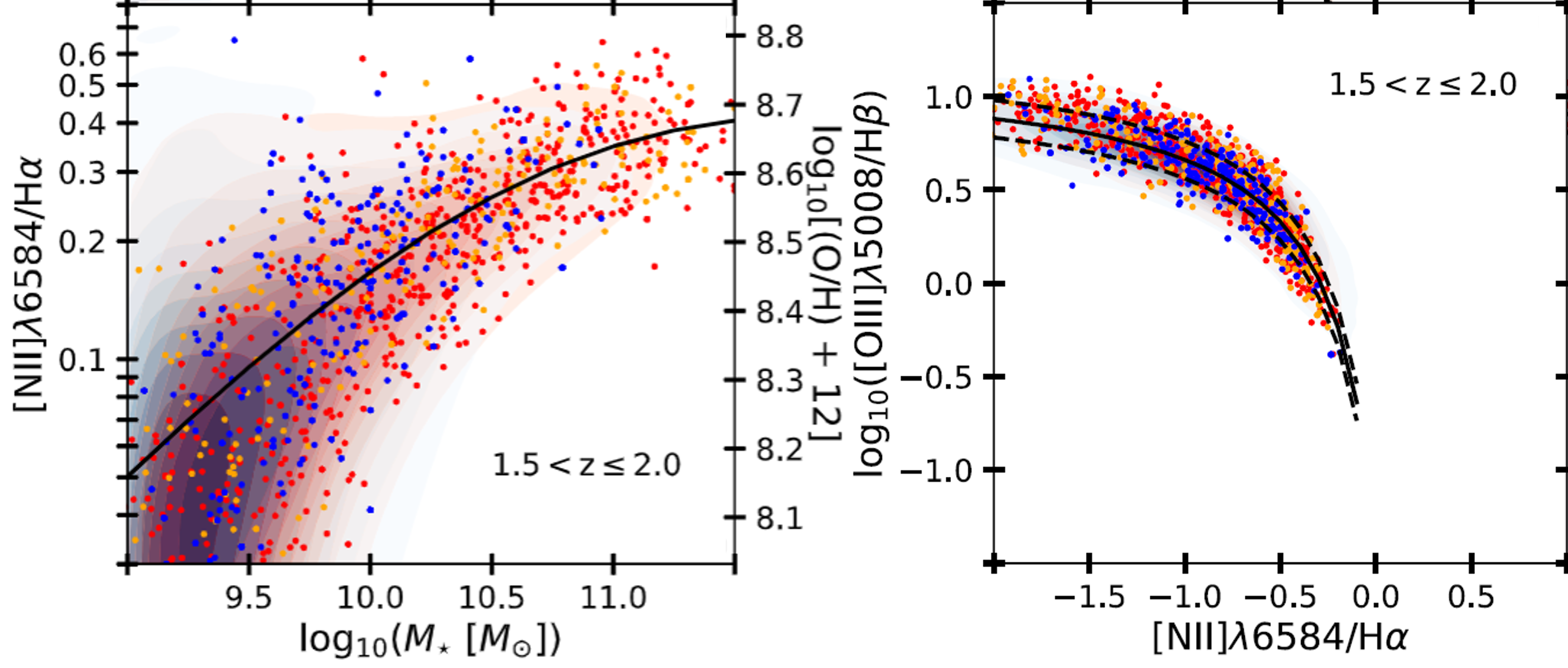}
    \caption{\emph{Left}: The $M_\star$-metallicity relation (MZR) as presented in \cite{Wuyts2014} in the redshift bin $1.5\leq z \leq 2.0$  with the predicted metallicity indicated on the right vertical axis. The left vertical axis indicates the corresponding ratio $N_2$ predicted using the linear calibrations presented in \cite{Asplund2004}. \emph{Right}: $N_2$-BPT diagram as presented in \cite{Kewley2013a} for sources between $1.5\leq z \leq 2.0$. See Euclid Collaboration: Gabarra et al., in prep, for further details.} \label{BPT}

\end{figure}
\subsection{Continuum reconstruction and emission lines flux predictions}

We reconstructed the continuum using the Bruzual \& Charlot (2003) models. We added the nebular emission Balmer, Paschen, [N\textsc{\lowercase{\,ii}}]$\lambda\lambda6584,6549$, [O\textsc{\lowercase{\,ii}}]$\lambda\lambda$3727,3729, [O\textsc{\lowercase{\,iii}}]$\lambda\lambda$5007,4959, [S\textsc{\lowercase{\,ii}}]$\lambda\lambda$6731,6717, and [S\textsc{\lowercase{\,iii}}]$\lambda\lambda$9531,9069 lines making use of calibrations available in the literature. We refer to empirical relations for emission line analysis such as H$\alpha$ luminosity-SFR \cite{Kennicutt1998}, the mass-metallicity relation, the Baldwin-Phillips-Terlevich diagram (BTP, see Fig. \ref{BPT}), and photoionisation models. The emission line fluxes are then transformed into Gaussian profiles and integrated to the continuum. 

The \Euclid spectroscopy will be performed in slitless configuration. This can be thought as if the \textit{slit} were the galaxy itself. A strong effect of the galaxy shape on the quality of the extracted spectra is therefore expected. The shape of the galaxy will mainly affect the signal to noise ratio and the full width at half maximum (FWHM) of the emission lines (see Euclid Collaboration: Gabarra et al., in prep, for details). This aspect led us to take a special care in modelling the shape of the emission lines, considering the velocity dispersion. To account for it, we applied the velocity dispersion to stellar mass relation proposed by Bezanson et al. (2018). 

\section{Results and conclusion}
\begin{figure}[h]
    \includegraphics[width=\textwidth]{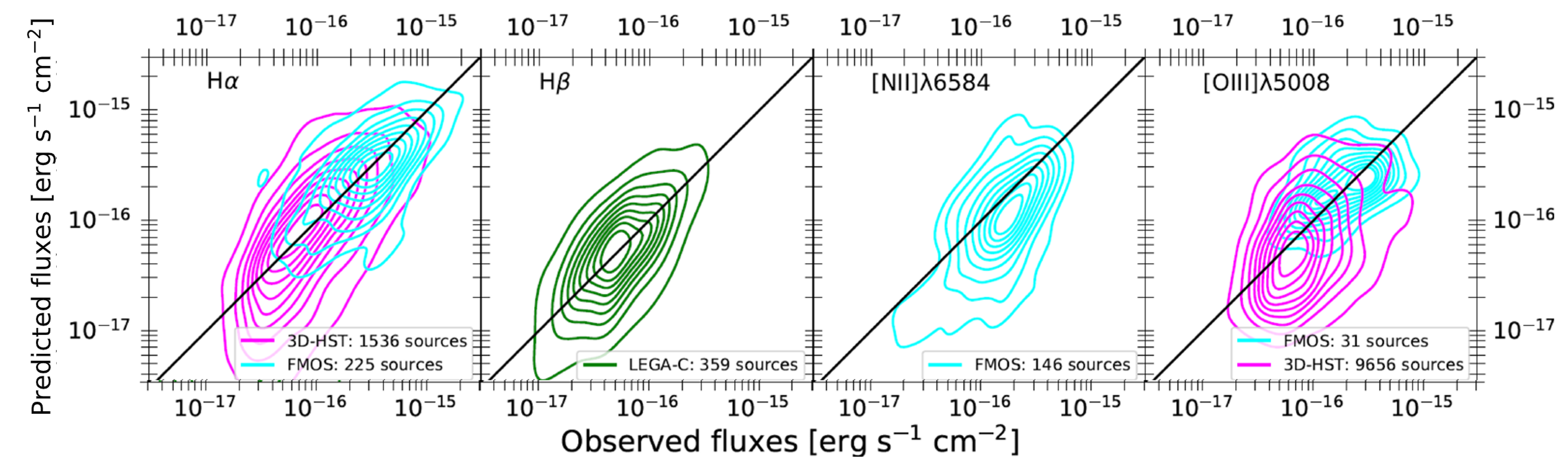}
      \caption{Comparison of the predicted fluxes to the observed fluxes coming from publicly released data of the near-infrared spectroscopic surveys
3D-HST (purple), FMOS (green) and LEGA-C (cyan). The contours correspond to iso-proportions of the distribution density starting at 20\% with
a 10\% step. See Euclid Collaboration: Gabarra et al., in prep, for further details.} \label{comparisonELS}
\end{figure}

\begin{figure}[h]  
\includegraphics[width=\textwidth]{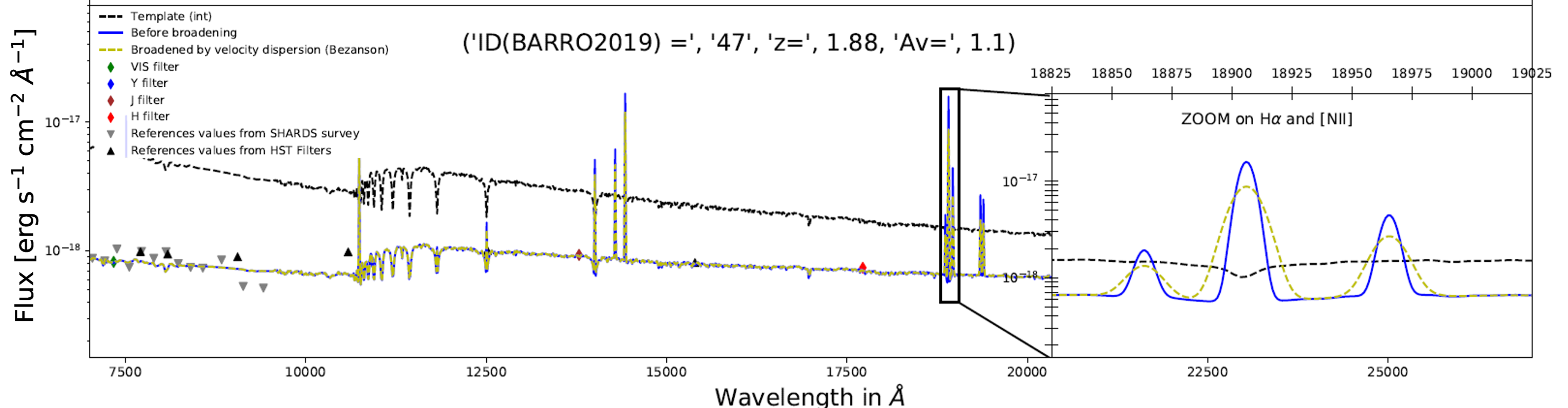}
  \caption{Examples of photometric evaluation of the incident spectra created in this work. The source is from the B19 catalogue and the continuum is compared to observational data from the SHARDS and HST surveys photometry (grey and black triangles). The zoom-in is made on the H$\alpha$-[N{\textsc{\lowercase{\,ii}}}]$\lambda\lambda$6584,6549 triplet. The coloured diamonds indicate the fluxes obtained with the \textit{Euclid} filters.}
  \label{comparison_continuum}
\end{figure}

Spectroscopic data are available in the B19 and L16 catalogues for some of the emission lines predicted in this work. We present in Fig. \ref{comparisonELS} a comparison between our predictions and observation. In Fig. \ref{comparison_continuum}, we present an example of a spectrum for a source from the B19 catalogue. We indicate on the figure the magnitudes  from observation within the SHARDS and HST surveys. 

We presented the construction of spectroscopic data starting from publicly released catalogues from the COSMOS and CANDELS/GOODS-N fields. The SEDs are reconstructed based on stellar population models using the software \texttt{GALAXEV}, and we calculated the nebular emission line fluxes using observed scaling relations and photoionisation models.\ The constructed SEDs have been compared and validated using photometric and spectroscopic data available in the catalogues. The good agreement between the predictions and the observations served as a validation test to attest for the realistic emission line flux estimations.

\footnotesize
\textit{Acknowledgements}. The author thanks Chiara Mancini, Lucia Rodríguez-Muñoz, Giulia Rodighiero, Chiara Sirignano, Marco Scodeggio and Margherita Talia for their help in shaping this work. \AckEC.
\setlength{\bibsep}{0pt plus 0.3ex}

\bibliography{bib}
\end{document}